\documentclass{article}
\usepackage[margin=0.95in]{geometry}
\usepackage[parfill]{parskip}
\usepackage[utf8]{inputenc}
\usepackage[super,sort&compress]{natbib}
\setcitestyle{citesep={,}}
\usepackage{amsmath,amssymb,amsfonts,amsthm}
\usepackage{array} 
\usepackage{url}
\usepackage{caption}
\usepackage{graphicx}
\usepackage[switch]{lineno}

\begin{document}

\let\numbercite\cite
\let\numbercitep\citep
\renewcommand{\cite}[1]{Aygün et al.\numbercite{#1}}
\renewcommand{\citep}[1]{\unskip\numbercitep{#1}}
\renewcommand{\citealt}[1]{\citep{#1}}

\title{Information leakage from data revisions in retrospective forecasts}
\author{Johannes Bracher$^{1,2,*}$ and Sebastian Funk$^{3,*}$\\[0.5em]
\small $^1$Institute of Statistics, Karlsruhe Institute of Technology, Germany\\
\small $^2$Computational Statistics Group, Heidelberg Institute for Theoretical Studies, Germany\\
\small $^3$Centre for Mathematical Modelling of Infectious Diseases, London School of Hygiene \& Tropical Medicine,\\
\small London, UK\\
\small $^*$Corresponding authors}
\date{}

\maketitle

\noindent\fbox{%
    \parbox{\textwidth}{%
        \textit{This manuscript presents a re-analysis and discussion of a case study by Ayg\"un et al (2026, Nature, \url{https://doi.org/10.1038/s41586-026-10658-6}). In case the authors of the original paper post a reply, we will link it here.}
    }%
}

\paragraph{Summary.} \cite{Ayguen2026} claim that their AI-driven \textit{Empirical Research Assistance (ERA)} system produces COVID-19 hospitalisation forecasts which outperform the state-of-the-art CDC ensemble by a considerable margin for the 2024/25 season. We demonstrate that the observed performance gain is attributable to information leakage in the retrospective forecasting setup, which resulted because data revisions were not taken into account. As similar mechanisms are at play in many other forecasting fields, our cautionary tale applies not just to epidemic forecasting, but is relevant to the entire emerging field of AI-assisted predictive modelling.

\paragraph{Background: collaborative infectious disease forecasting.} \textit{Forecast Hubs} are widely used to structure predictive modelling activities in infectious disease epidemiology \citep{Reich2022}. These platforms collect forecasts of relevant indicators by different research teams according to standardised schedules and formats. The Hubs facilitate principled comparisons of different forecasting approaches and their combination into ensemble predictions. Besides their increasing integration into the work of public health agencies \citep{cramer2022, sherratt2022}, they serve as benchmarking systems for new methods such as \textit{Inferno}\citep{Osthus2022}. Their easily understandable prediction task, reliance on quantitative evaluation metrics and their public display of leader boards have recently made them a popular case study for AI-driven scientific modelling and programming.\citep{Dudley2026, Lu2026, Wang2026}

\paragraph{Data revisions and information leakage.} In this note we point out a major pitfall of retrospectively conducted benchmarking on the Forecast Hubs: real-time disease surveillance data are often heavily revised due to reclassification or reports arriving at different delays. Analyses built on the consolidated data available at the end of the season can therefore leak information that was not available when the forecasts collated in the Hubs would have been issued. The challenge of data revisions is well-known \citep{Reich2019,  cramer2022, sherratt2022,Mathis2024}, and dedicated systems exist to retrieve historic data snapshots \citep{Reinhart2021}. Nonetheless, retrospective studies building on the Hub archives regularly fail to appropriately reconstruct the data as they were available at the time the Hub forecasts were collated. Many authors seem to underestimate the resulting biases, and merely discuss the use of consolidated data as a limitation \citep{Wang2026}. We document how large these biases can be, and argue that benchmarking in the Hub infrastructure is best done prospectively. Whenever analyses are done retrospectively, it is crucial to carefully reconstruct real-time data, for which dedicated tooling that fetches and freezes vintaged ground truth for reproducible scoring is beginning to emerge.\citep{EpiBenchmark} Similar challenges arise in other forecasting fields, as discussed by Croushore\citep{Croushore2006} for macroeconomic predictions.

\paragraph{COVID-19 case study by \cite{Ayguen2026}.} We illustrate the problem with the COVID-19 case study of \cite{Ayguen2026}. Its prominence and publication in \emph{Nature} make it likely to inspire similar applications by others, and thus a particularly relevant example. It also lets us make the point without casting doubt on the proposed method itself, as the authors have since demonstrated convincing real-time performance in a genuinely prospective setting \citep{Martinson2026}.

\cite{Ayguen2026} present \textit{ERA}, which combines large language models and tree search to generate empirical software in settings with a quantitative target metric. One of their six case studies addresses retrospective forecasting of COVID-19 hospitalisations during the 2024/25 season. Predictions were generated for 52 US states and territories for lead times of 0 to 3 weeks. In what is described as a ``rigorous retrospective study to assess ERA performance in this competitive environment'' (p6), the authors claim to achieve a substantial performance improvement over the state-of-the-art CDC Forecast Hub ensemble, evidenced by a reduction in mean Weighted Interval Score~(WIS) of 11\% (from 29.1 to 25.9).

\begin{figure}[h!]
    \centering
    \includegraphics[width=0.8\linewidth]{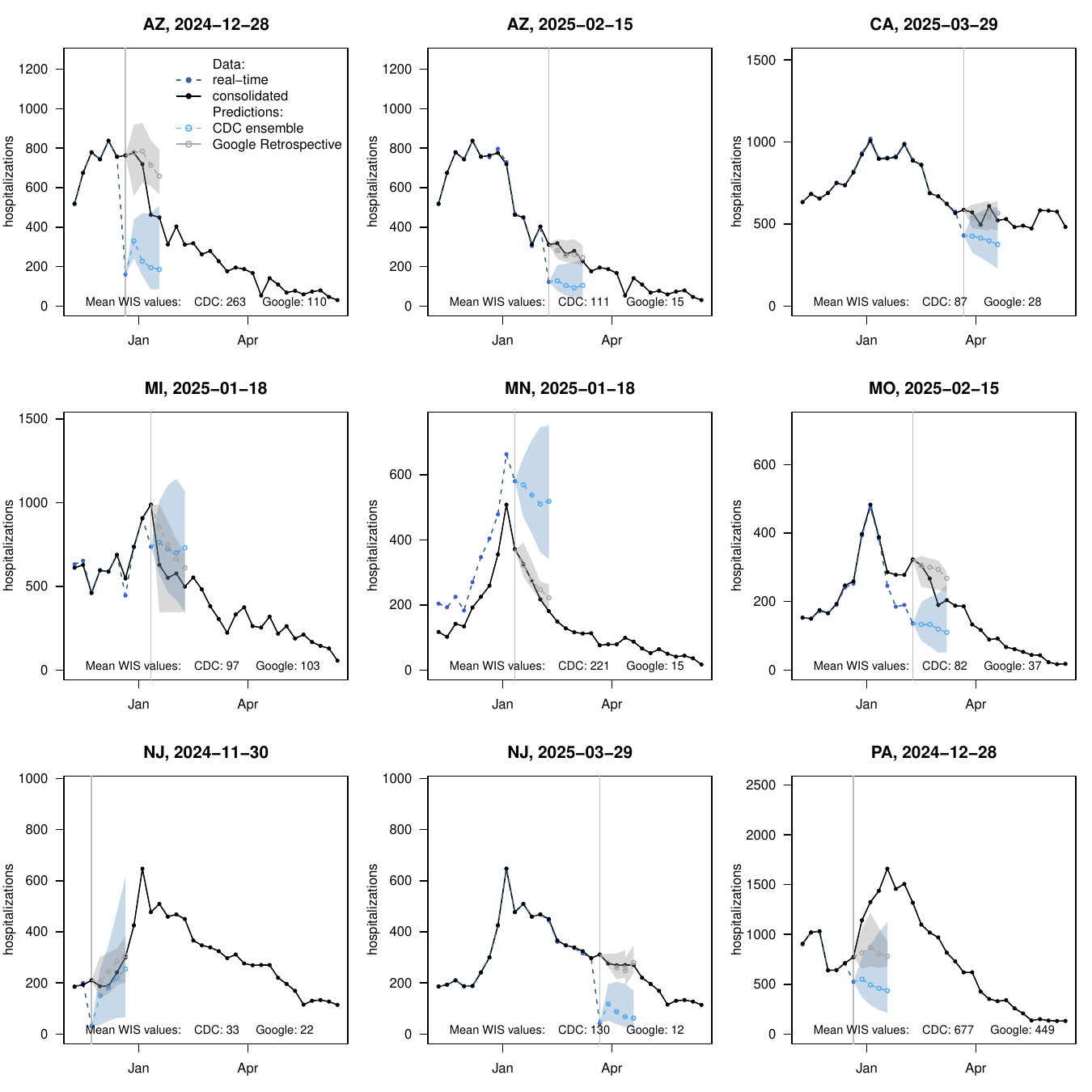}
    \caption{Illustration of the nine state / reference week combinations with the most pronounced data revisions (in terms of absolute difference for the last available data point at the time of forecasting). The figure overlays real-time and consolidated data, as well as forecasts by \textit{Google Retrospective} and the CDC ensemble. Differences in mean WIS across the four forecast horizons are shown at the bottom of each panel.}
    \label{fig:extreme_cases}
\end{figure}

\paragraph{Assessing the role of data revisions.} The authors argue that the ``improvements in WIS are driven by superior predictive accuracy across diverse temporal regimes, rather than statistical artifacts'' (Supplement, p12). We disagree with this conclusion, as virtually the entire performance gain can be explained by information leakage, resulting from a failure to take data revisions into account. Some hints in this direction may be found in the results presented by \cite{Ayguen2026}. Firstly, as the holiday season is especially prone to data revisions, the vast improvement for reference date 28 December 2024 warrants scrutiny (mean WIS reduced from 73 to 49, accounting for 28\% of the total performance difference). Secondly, Supplementary Figures 7 and 8 show signatures of data revisions, especially for Washington DC and Puerto Rico. The CDC ensemble forecasts are repeatedly misaligned with the preceding consolidated data points, suggesting discrepancies with real-time data.

Individually, these signals are only suggestive, but the authors shared replication materials upon our request. In Figure \ref{fig:extreme_cases} we display forecasts from their \textit{Google Retrospective} approach and the CDC ensemble for the nine instances with the strongest revisions of the last available data point (in terms of absolute differences). The CDC ensemble was clearly led astray by unreliable real-time data in these cases, while \textit{Google Retrospective} could use consolidated data. These nine instances (out of 1430) account for 20\% of the total reported WIS difference, with mean WIS values differing by more than a factor of two (88 for \textit{Google Retrospective} vs. 189 for the CDC Ensemble).

Table \ref{tab:table} provides a more comprehensive assessment. We divided the evaluation set into five roughly equally-sized bins according to how strongly the last data point available at the time of forecasting was ultimately revised, once by relative (top) and once by absolute revisions (bottom). Performance is summarised by relative WIS values (ratios of mean WIS for \textit{Google Retrospective} and the CDC Ensemble), computed per bin and cumulatively for each bin and all bins below. In agreement with the results reported by \cite{Ayguen2026}, we obtain a relative WIS of 0.89 across the entire evaluation set (11\% improvement; bottom right cell of each sub-table). However, it is clearly visible that performance gains are concentrated in bins corresponding to strong relative or absolute revisions. The strongest relative improvement is achieved for state-weeks where the last data point was revised by a quarter or more (relative WIS 0.69 / 31\% improvement). In the bins corresponding to no or minimal revisions (at most 1\% or an absolute difference of 1), the two models are almost exactly on par. We hence conclude that the improvements reported by \cite{Ayguen2026} are driven by information leakage. While performance matching the CDC ensemble remains impressive, it is less spectacular than the reported 11\% improvement.

\begin{table}[ht]
    \centering
\begin{tabular}{rrrrrr}
\hline
\multicolumn{6}{c}{\textbf{Binning according to relative revisions}} \smallskip \\
  \hline
 Bin & [0\%, 1\%] & (1\%, 5\%] & (5\%, 10\%] & (10\%, 25\%] & (25\%, 380\%] \\ 
  \hline
Fraction of state-weeks & 20\% & 21\% & 20\% & 22\% & 18\% \\ 
  Rel. WIS in this bin & 1.00 & 0.92 & 0.95 & 0.90 & 0.69 \\ 
  Rel. WIS in this bin and bins below & 1.00 & 0.94 & 0.94 & 0.93 & 0.89 \\ 
   \hline \\ \hline
   \multicolumn{6}{c}{\textbf{Binning according to absolute revisions}} \smallskip\\
   \hline
    & [0, 1] & (1, 5] & (5, 10] & (10, 25] & (25, 603] \\ 
  \hline
Fraction of state-weeks & 27\% & 21\% & 13\% & 20\% & 19\% \\ 
  Rel. WIS in this bin & 1.01 & 0.95 & 0.99 & 0.91 & 0.81 \\ 
  Rel. WIS in this bin and bins below & 1.01 & 0.98 & 0.98 & 0.95 & 0.89 \\ 
   \hline
\end{tabular}
    \caption{Relative WIS values of \textit{Google Retrospective} versus the CDC ensemble, i.e., $\overline{\text{WIS}}_\text{Google} / \overline{\text{WIS}}_\text{CDC}$ in subsets of the evaluation period. Top: Binning according to relative changes in the last data point available at the time of prediction. Bottom: Binning according to the corresponding absolute differences.}
    \label{tab:table}
\end{table}

\paragraph{Real-time forecasting performance of ERA.} In \cite{Ayguen2026}, the authors note that they ``ignor[e] potential differences in data available at the date of forecast'' (p23), but they likely underestimated their relevance. Since the completion of the present manuscript, a follow-up study was conducted in which a revised version of \textit{Google Retrospective} called \textit{Google SAI Ensemble} competed under real-time conditions during the 2025/26 season \citep{Martinson2026}. By submitting forecasts to the relevant Hub platforms in real time, the study avoided the issue of information leakage, and the model was found to top the leader board for all three diseases considered. We note that this is not at odds with our findings for the 2024/25 season, both because the ERA methodology was substantially revised (now using an ensemble rather than model selection) and because relative performance can fluctuate markedly over time.

\paragraph{Data and code availability.} The data underlying the presented analyses have kindly been made available by \cite{Ayguen2026} at \url{https://github.com/google-research/era/tree/main/data}. Analysis codes to reproduce Figure \ref{fig:extreme_cases} and Table \ref{tab:table} can be found in the repository \url{https://github.com/jbracher/era-replication}.

\paragraph{Author contributions.} J.B. and S.F. jointly conceptualized the study, developed the methodology, conducted the investigation, interpreted the results and wrote and revised the manuscript.

\paragraph{Competing interests.} The authors declare no competing interests.

\paragraph{Additional information.} Correspondence and requests for materials should be addressed to J.B. (johannes.bracher@kit.edu) or S.F. (sebastian.funk@lshtm.ac.uk).

\end{document}